\documentclass[conference]{IEEEtran}
\IEEEoverridecommandlockouts

\usepackage{cite}
\usepackage{amsmath,amssymb,amsfonts}
\usepackage{algorithmic}
\usepackage{graphicx}
\usepackage{textcomp}
\usepackage{xcolor}
\usepackage{booktabs}
\usepackage{url}
\usepackage{multirow}
\usepackage{colortbl}
\usepackage{xparse}
\def\BibTeX{{\rm B\kern-.05em{\sc i\kern-.025em b}\kern-.08em
    T\kern-.1667em\lower.7ex\hbox{E}\kern-.125emX}}

\newcommand{\Transfers}{\mathcal{T}}   
\newcommand{\XLinks}{\mathcal{L}}      
\newcommand{\Chains}{\mathcal{C}}      
\newcommand{\Assets}{\mathcal{A}}      
\newcommand{\Bridges}{\mathcal{B}}     

\newcommand{\Spender}{\mathsf{Spender}}
\newcommand{\Recipient}{\mathsf{Recipient}}
\newcommand{\transfer}{\tau}            
\newcommand{\xlink}{\ell}               
\newcommand{\chain}{\mathsf{chain}}     
\newcommand{\asset}{\mathsf{asset}}     
\newcommand{\ts}{\mathsf{time}}         
\newcommand{\amt}{\mathsf{amount}}      
\newcommand{\src}{\mathsf{src}}          
\newcommand{\dst}{\mathsf{dst}}          

\newcommand{\Pred}{\mathsf{Pred}}
\newcommand{\Anc}{\mathsf{Anc}}

\definecolor{lightgray}{gray}{0.93}

\begin{document}

\title{LOCARD: An Agentic Framework for Blockchain Forensics}


\author{\IEEEauthorblockN{1\textsuperscript{st} Xiaohang Yu}
\IEEEauthorblockA{\textit{Imperial College London} \\
x.yu21@imperial.ac.uk}
\and
\IEEEauthorblockN{2\textsuperscript{nd} William Knottenbelt}
\IEEEauthorblockA{\textit{Imperial College London} \\
w.knottenbelt@imperial.ac.uk}
}

\maketitle

\begin{abstract}
Blockchain forensics inherently involves dynamic and iterative investigations, while many existing approaches primarily model it through static inference pipelines.
We propose a paradigm shift towards Agentic Blockchain Forensics (ABF), modeling forensic investigation as a sequential decision-making process.
To instantiate this paradigm, we introduce LOCARD, the first agentic framework for blockchain forensics.
LOCARD operationalizes this perspective through a Tri-Core Cognitive Architecture that decouples strategic planning, operational execution, and evaluative validation. 
Unlike generic LLM-based agents, it incorporates a Structured Belief State mechanism to enforce forensic rigor and guide exploration under explicit state constraints.
To demonstrate the efficacy of the ABF paradigm, we apply LOCARD to the inherently complex domain of cross-chain transaction tracing.
We introduce Thor25, a benchmark dataset comprising over 151k real-world cross-chain forensic records, and evaluate LOCARD on the Group-Transfer Tracing task for dismantling Sybil clusters.
Validated against representative laundering sub-flows from the Bybit hack, LOCARD achieves high-fidelity tracing results, providing empirical evidence that modeling blockchain forensics as an autonomous agentic task is both viable and effective.
These results establish a concrete foundation for future agentic approaches to large-scale blockchain forensic analysis.
Code and dataset are publicly available at \url{https://github.com/xhyumiracle/locard} and \url{https://github.com/xhyumiracle/thorchain-crosschain-data}.
\end{abstract}


\section{Introduction}

The transparency and immutability of blockchain ledgers have made transaction data a valuable source for forensic investigation, enabling the tracing and attribution of illicit fund flows for purposes ranging from criminal investigation to regulatory compliance. 
However, the rapid proliferation of decentralized finance (DeFi) protocols and cross-chain interoperability mechanisms has fundamentally reshaped the threat landscape. 
Illicit actors increasingly exploit chain hopping, decentralized bridges, and complex transaction compositions to obfuscate money flows, transforming blockchain forensics from a relatively localized tracing problem into a large-scale, adversarial investigation spanning heterogeneous ledgers and asset semantics~\cite{chainalysis_2024, electronics13173568,kumar2025surveytransactiontracingtechniques}.

\paragraph{Blockchain forensics and transaction tracing.}
Prior research in blockchain forensics has developed a rich set of techniques for tracing transactions and analyzing behavioral patterns on individual blockchains. 
Representative systems model the ledger as a transaction or account graph and apply graph traversal, temporal partitioning, or heuristic propagation to follow the flow of funds~\cite{tracer_2024, tpgraph_2024, lin2023understandingcryptomoneylaundering, sheddingLightonShadows}. 
Complementary efforts have produced general-purpose forensic analytics platforms, such as GraphSense, which support large-scale exploration and investigation of cryptocurrency transaction data~\cite{graphsense_haslhofer2021graphsensegeneralpurposecryptoassetanalytics}.
In parallel, recent work has also explored large language models (LLM) for blockchain analysis tasks, such as transaction representation learning and anomaly detection~\cite{gai2023blockchainlargelanguagemodels}.
While effective within a single ledger, these approaches largely frame tracing as a static analysis pipeline, assuming that the investigative logic can be predefined and executed in a fixed manner.

\begin{figure}
    \centering
    \includegraphics[width=\linewidth]{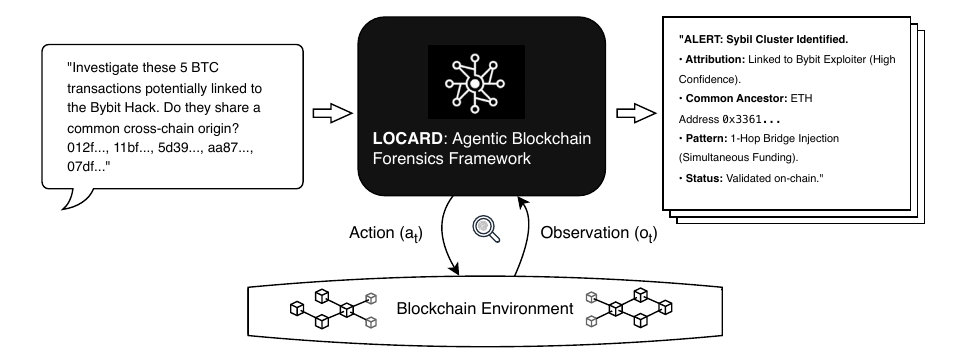}
    \caption{LOCARD: Agentic Blockchain Forensics Framework}
    \label{fig:locard}
\end{figure}

\paragraph{Cross-chain forensics and emerging challenges}
As assets increasingly move across blockchains through centralized exchanges and decentralized bridge protocols, recent work has begun to address cross-chain transaction tracing. 
Early efforts associate transactions across ledgers via exchange-mediated address clustering or rule-based linkage~\cite{cltracer_2022, lin2024connectorenhancingtraceabilitydecentralized}. 
More recent systems leverage event-log mining and learning-based association to automatically uncover cross-chain transfer relations across multiple bridge protocols~\cite{piecingTogethertheJigsawPuzzle, abctracer}. 
Parallel lines of work focus on detecting abnormal cross-chain behaviors or accounts using graph-based or multi-model learning techniques~\cite{crosschainabnormaltransactiondetection}. 
Recent efforts have also focused on constructing large-scale cross-chain datasets and measurement frameworks to support forensic analysis across heterogeneous ledgers~\cite{piecingTogethertheJigsawPuzzle, augusto2025xchaindatagencrosschaindatasetgeneration, anempiricalstudyoncrosschain,  graphsense_haslhofer2021graphsensegeneralpurposecryptoassetanalytics}.
Despite these advances, cross-chain transaction tracing remains an inherently complex forensic task: it spans heterogeneous ledgers with no explicit on-chain linkage, induces large combinatorial candidate spaces, and must operate under evolving adversarial strategies.

\paragraph{Limitations of static forensic pipelines}
A common characteristic across existing tracing systems is that investigative logic is encoded as static pipelines, fixed heuristics, or task-specific association rules. 
Such designs are effective when laundering patterns are stable, but they lack the ability to adapt, revise hypotheses, or strategically explore alternative explanations as new evidence emerges. 

\paragraph{Toward agentic blockchain forensics}
In contrast, real-world forensic investigation is inherently iterative and strategic: investigators form hypotheses, gather evidence, validate consistency, and backtrack when confronted with contradictions.
Recent advances in agentic systems and large language models have introduced a range of cognitive and control abstractions for modeling iterative problem-solving processes, including sense--think--act cycles, belief--desire--intention (BDI) models~\cite{rao1995bdi}, and reasoning--action frameworks such as ReAct~\cite{yao2023reactsynergizingreasoningacting}.
These agentic abstractions are commonly understood through the lens of sequential decision-making, with formalisms such as partially observable Markov decision processes (POMDPs) providing a principled framework for reasoning in dynamic environments~\cite{zhang2025landscapeagenticreinforcementlearning}.
However, despite their success in other domains, agentic formulations remain largely unexplored in blockchain forensics, particularly for high-complexity tasks such as cross-chain transaction tracing.

\paragraph{The exploration-exploitation tension}
Blockchain forensic investigation is subject to two opposing pressures: adversarial obfuscation demands exploration of new hypotheses, while forensic verifiability demands exploitation of established findings. Static pipelines collapse onto exploitation: rigorous but inflexible, they cannot adapt as adversaries evolve. LLM-based agents, by contrast, collapse onto exploration: flexible but undisciplined, they cannot maintain evidential rigor. \textit{How to architect an agentic system that is both flexible and rigorous in blockchain forensics} is therefore a hard, open problem.

To address these challenges, we advocate for a paradigm shift towards \textit{Agentic Blockchain Forensics (ABF)}. Unlike static heuristics, ABF conceptualizes forensic investigation as a dynamic, sequential decision-making process. 
In this work, we present LOCARD\footnote{We name the system LOCARD in honor of Edmond Locard, a pioneer of forensic science, and his fundamental principle: ``Every contact leaves a trace'', which profoundly resonates with the immutable nature of blockchain ledgers.}, a framework designed to operationalize this paradigm (Figure~\ref{fig:locard}). We instantiate the system within the high-complexity domain of cross-chain tracing to validate its efficacy.
Our specific contributions are summarized as follows:

\begin{itemize}
    \item {Paradigm:} We propose Agentic Blockchain Forensics (ABF), a paradigm that models blockchain forensic investigation as a sequential decision-making process, an iterative and adaptive investigation unfolding through continuous interaction with the blockchain environment, rather than a static retrieval or inference task.
    
    \item {Framework:} We present LOCARD, the first ABF framework.
    Its \textit{Tri-Core Cognitive} Architecture decouples strategic planning, operational execution, and evaluative validation. Its \textit{Structured Belief State} ensures rigorous reasoning over complex evidence trails. Together they operationalize the balance between exploration and exploitation, addressing the long-horizon and logic-sensitive nature of forensic tasks.
    
    \item {Dataset:} We introduce and open-source \textit{Thor25}, a comprehensive real-world cross-chain benchmark. It comprises over \textit{151.5k} ground-truth records spanning four major Layer-1 blockchains (BTC, ETH, DOGE, LTC) throughout 2025. This dataset serves as a critical resource for evaluating cross-chain tracing algorithms in realistic environments.
    
    \item {Application:} We apply the system to the advanced Group-Transfer Tracing task. By reconstructing representative laundering sub-flows within the real-world \textit{Bybit Hack} incident, we demonstrate LOCARD's capability to dismantle Sybil Clusters and uncover shared entities behind disjoint illicit flows.
\end{itemize}

\section{Problem Modeling}

\subsection{Preliminaries}
\label{sec:preliminaries}

\paragraph{Chains, assets and bridges.}
Let $\Chains$ denote a set of blockchains, each supporting a native asset.
Let $\Assets$ denote a set of assets, where an asset may be supported by multiple blockchains,
and each blockchain $c \in \Chains$ supports a subset of assets from $\Assets$.
Let $\Bridges$ denote a set of cross-chain infrastructures (``bridges'').

\paragraph{On-chain transfers.}
Let $\Transfers$ denote the set of observable on-chain transfer events.
Each transfer $\transfer \in \Transfers$ corresponds to a blockchain transfer event and is characterized by 
\[
\transfer
\;\triangleq\;
\begin{aligned}
\bigl(
&\chain(\transfer),
 \ts(\transfer),
 \asset(\transfer),\\
&\amt(\transfer),
 \Spender(\transfer),
 \Recipient(\transfer)
\bigr)
\end{aligned}
\]

where $\chain(\transfer) \in \Chains$ denotes the underlying blockchain,
$\ts(\transfer)$ the timestamp,
$\asset(\transfer)$ the transferred asset,
$\amt(\transfer)$ the transferred amount,
and $\Spender(\transfer)$ and $\Recipient(\transfer)$ the sender-side and recipient-side address sets, respectively.
We treat a transfer as an atomic value-carrying event

\paragraph{Same-chain predecessor relation.}
On a given blockchain, we define a same-chain predecessor relation
\[
\tau' \prec \tau,
\]
which indicates that transfer $\tau'$ is a traceable value-flow predecessor of
transfer $\tau$ on the same chain.
The relation $\prec$ is instantiated by the transaction semantics of the
underlying chain (e.g., spent-output dependencies for UTXO-based chains and
sender-to-recipient dependencies for account-based chains).

By definition, $\prec$ respects the canonical chain order \footnote{$\mathrm{ord}(\cdot)$ denotes the total order induced by the blockchain, including both block height and intra-block transaction order.}:
\[
\tau' \prec \tau \;\Rightarrow\; \mathrm{ord}(\tau') \le \mathrm{ord}(\tau),
\]

We write 
\[
\tau' \prec^{\le H} \tau,
\]
to denote predecessor reachability within at most $H$ hops.

\paragraph{Cross-chain transfer links.}

A cross-chain transfer link is represented as a tuple
\[
\xlink = (\transfer^{\mathrm{src}}, \transfer^{\mathrm{dst}}, b),
\]
where $\transfer^{\mathrm{src}} \in \Transfers$ is the source-side transfer,
$\transfer^{\mathrm{dst}} \in \Transfers$ is the destination-side transfer,
and $b \in \Bridges$ specifies the execution semantics under which the association is established.
In most cases, $\chain(\transfer^{\mathrm{src}}) \neq \chain(\transfer^{\mathrm{dst}})$.

More generally, a cross-chain link represents an association across execution domains
that cannot be recovered from single-chain transaction semantics alone.

Let $\XLinks$ denote the set of all such cross-chain transfer links.

\begin{figure}
    \centering
    \includegraphics[width=\columnwidth]{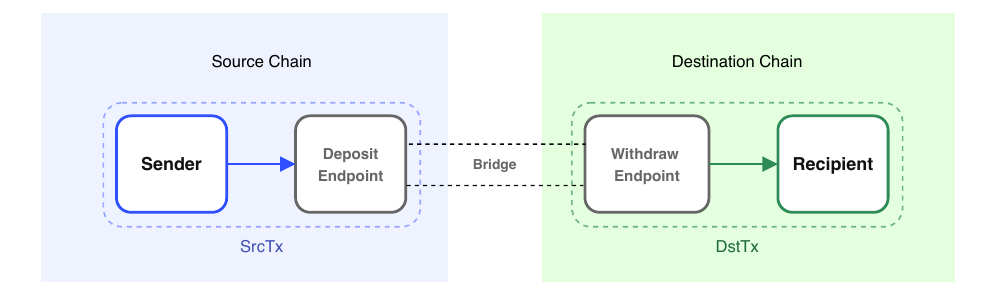}
    \caption{Cross-chain Transaction through a Bridge}
    \label{fig:single-transfer-demo}
\end{figure}

\subsection{Cross-chain Tracing Tasks}
\label{sec:tasks}

We formalize cross-chain tracing as the problem of recovering ground-truth
cross-chain links between on-chain transfer events.
We consider two tracing tasks: single-transfer tracing and group-transfer tracing.

\subsubsection{Single-Transfer Tracing}
\label{sec:single-task}

Given a transfer $\tau^\ast \in \Transfers$,
single-transfer tracing aims to recover all ground-truth cross-chain links
whose destination-side transfer is $\tau^\ast$ (Figure~\ref{fig:single-transfer-demo}).

Formally, the ground-truth solution is defined as
\[
\mathcal{L}_{\mathrm{gt}}(\tau^\ast)
\;=\;
\{\, \ell \in \XLinks \mid \dst(\ell) = \tau^\ast \,\}.
\]
The objective of single-transfer tracing is to recover the set
$\mathcal{L}_{\mathrm{gt}}(\tau^\ast)$.

\subsubsection{Group-Transfer Tracing}
\label{sec:group-task}

Given a set of transfers
\[
\mathcal{Q} = \{\tau_1, \dots, \tau_k\} \subset \Transfers ,
\]
group-transfer tracing aims to recover all ground-truth cross-chain links
associated with the transfers in $\mathcal{Q}$ jointly.

In forensic contexts, such a set $\mathcal{Q}$ often represents a suspected \textit{Sybil Cluster}, i.e. a group of distinct addresses controlled by a single entity to obfuscate illicit fund flows through parallel execution.

Formally, the ground-truth solution is defined as
\[
\mathcal{L}_{\mathrm{gt}}(\mathcal{Q})
\;=\;
\{\, \ell \in \XLinks \mid \dst(\ell) \in \mathcal{Q} \,\}.
\]

Unlike single-transfer tracing, group-transfer tracing requires joint reasoning over $\mathcal{Q}$, because ground-truth links associated with different transfers may be correlated through shared same-chain upstream value-flow structure, which induces overlapping ancestry among their source-side transfers.

\subsection{Agentic Blockchain Forensics Paradigm}
\label{sec:abf}

We introduce \emph{Agentic Blockchain Forensics (ABF)} as a paradigm for modeling blockchain forensic investigation as a sequential decision-making process, rather than a static retrieval or inference task.
In ABF, forensic analysis is viewed as an iterative and adaptive investigation that unfolds through continuous interaction with the blockchain environment.

Under this paradigm, the investigation process is conceptualized as being driven by an autonomous investigative logic that incrementally reduces epistemic uncertainty over forensic hypotheses, such as transaction linkages, entity associations, and cross-chain relationships.
Rather than executing a predetermined sequence of queries, the investigation progressively gathers evidence, evaluates its consistency, updates its internal belief representation, and adapts subsequent investigative steps based on the evolving context.

\paragraph{Process Definition}
Formally, the agentic forensic investigation process is characterized as a tuple
\[
\langle \mathcal{S}, \mathcal{A}, \mathcal{O}, \Phi \rangle ,
\]
which captures the evolution of investigative state through admissible actions, environment observations, and belief updates.
Each component of this formulation is elaborated in the following paragraphs.

\paragraph{State Space}
An investigative state $S_t \in \mathcal{S}$ summarizes the current forensic context at step $t$.
It encodes observed evidences, unresolved uncertainties, and intermediate hypotheses accumulated over the course of the investigation.
The state is updated incrementally as new evidence is obtained and assessed.

\paragraph{Action Space}
The action space $\mathcal{A}$ captures the set of admissible investigative decisions that guide how the forensic investigation progresses.
In practice, actions may involve expanding the investigative scope, acquiring new evidence, assessing existing findings, or consolidating competing hypotheses.
This abstraction allows the investigation to proceed adaptively, without imposing a fixed procedural order or committing to a predefined investigation strategy.

\paragraph{Observation Space}
An investigative action $a_t \in \mathcal{A}$ results in an observation
$o_t \in \mathcal{O}$, which provides new information relevant to the ongoing forensic investigation.
Observations may include newly retrieved evidence, evaluative feedback or rationales of failure.

\paragraph{State Update}
The investigative state is updated by integrating newly acquired observations and executed actions.
Formally, the state evolves according to
\[
S_{t+1} = \Phi(S_t, a_t, o_t),
\]
where the state update function $\Phi$ specifies how new information is incorporated to refine the current investigative understanding through agentic reasoning.

\section{Cross-Chain Tracing Heuristics}
\label{sec:heuristics}

\subsection{Single-Transfer Tracing Heuristics}
\label{sec:heu-single-transfer}

We focus on the \emph{1-to-1} single-transfer cross-chain setting, where a single source-chain transfer gives rise to a single destination-chain transfer.
Given a destination transfer $\transfer^{dst}$ observed at time $t_d=\ts(\transfer^{dst})$ with transferred amount $A_d=\amt(\transfer^{dst})$, the objective is to infer plausible source transactions that could have causally produced it.

Although the heuristics are presented in a backward tracing setting, i.e. from observed destination transactions to potential sources, the same principles can be symmetrically applied to forward tracing. For clarity, we focus on backward tracing throughout this work.

\paragraph{Temporal constraint.}
Cross-chain transfers induce a directional temporal dependency from the source chain to the destination chain.
Accordingly, any valid source transfer must satisfy
\[
t_s \in \mathcal{T}_a
\;\triangleq\;
[\,t_d - \Delta t - \delta,\; t_d - \delta\,],
\]
where $\Delta t$ denotes the backward search window and $\delta \ge 0$ captures potential execution or settlement delays introduced by cross-chain systems.
In practice, this temporal window may be slightly relaxed to account for block timestamp imprecision across chains.
This constraint substantially reduces the candidate space, particularly on high-throughput chains.

\paragraph{Value-bounded backward search.}
Conditioned on the temporal window $\mathcal{T}_a$, let $P(t)$ denote the exchange rate between the destination asset and the source asset for $t \in \mathcal{T}_a$.
A feasible price range is derived as
\[
[P_{\min}, P_{\max}]
=
\bigl[
\min_{t \in \mathcal{T}_a} P(t),\;
\max_{t \in \mathcal{T}_a} P(t)
\bigr],
\]
from which a corresponding source-value interval is obtained:
\[
\mathcal{V}_s
=
A_d \cdot [P_{\min}, P_{\max}] \cdot [1 - \epsilon_p,\; 1 + \epsilon_p],
\]
where $\epsilon_p$ is a configurable buffer accounting for short-term volatility, execution delays, and heterogeneous price sourcing.
Only source transfers whose transferred amount $A_s$ satisfies
\[
A_s \in \mathcal{V}_s
\]
are retained as candidates.

\paragraph{Value-consistency forward validation.}
\label{sec:heu-single-validation}
For each retained candidate source transfer occurring at time $t_s \in \mathcal{T}_a$, a forward value consistency check is applied by evaluating the implied destination value using a tighter price range centered at $t_s$.
A candidate is considered valid only if the effective cross-chain fee is non-negative:
\begin{equation}
\label{eq:heu-value-validation}
    A_s \cdot P_{\max}(t_s) - A_d \ge 0
\end{equation}

This forward validation eliminates economically inconsistent candidates while remaining robust to transient price fluctuations and bridge-specific pricing mechanisms.

\subsection{Group Transfer Tracing Heuristics}
\label{sec:heu-group-transfer}

While single-transfer tracing analyzes each cross-chain transfer in isolation,
multiple transfers may originate from shared funding activity on the source chain.
Group transfer tracing exploits such shared structure by aggregating same-chain
upstream evidence across multiple transfers.

\paragraph{Same-chain upstream tracing.}
Given a transfer $\transfer$, we define its same-chain predecessor transfer set
using the predecessor relation $\prec$ as
\[
\Pred_H(\transfer)
\;\triangleq\;
\{\, \tau' \mid \tau' \prec^{\le H} \transfer \,\},
\]
where $H$ bounds the upstream tracing depth.
In practice, the predecessor set is further heuristically constrained to favor
proximate and meaningful funding activity (e.g., bounded expansion and exclusion of
negligible-value transfers).

The corresponding ancestor spender set is defined as
\[
\Anc(\transfer)
\;\triangleq\;
\bigcup_{\tau' \in \Pred_H(\transfer)} \Spender(\tau').
\]
which collects spender-side addresses involved in upstream value-giving events.

\paragraph{Address-level co-occurrence voting.}
Let $\Transfers^{\dst} = \{\transfer^{\dst}_1, \ldots, \transfer^{\dst}_N\}$ denote a
set of destination transfers under consideration, and let $\XLinks$ denote the set
of cross-chain link candidates produced by the single-transfer tracing process.
For each destination transfer $\transfer^{\dst}_i$, we define the corresponding
candidate source-side transfer set as
\[
\mathcal{S}(\transfer^{\dst}_i)
\;\triangleq\;
\{\, \transfer^{\src} \mid (\transfer^{\src}, \transfer^{\dst}_i, b) \in \XLinks \,\},
\]
which may be empty or contain multiple elements.

Group-level evidence is aggregated at the destination-transfer level.
Each destination transfer $\transfer^{\dst}_i$ casts at most one vote for an address
$a$ if $a$ appears in the ancestor spender set of \emph{any} of its source candidates,
The resulting hit count for an ancestor candiddate address $a$ is defined as
\[
\mathrm{hit}(a)
\;\triangleq\;
\sum_{i=1}^{N}
\mathbf{1}
\!\left(
a \in \bigcup_{\transfer \in \mathcal{S}(\transfer^{\dst}_i)} \Anc(\transfer)
\right),
\]
which measures how many destination transfers admit $a$ as a plausible common
upstream spender across their source-side candidates.

\paragraph{Common ancestor identification.}
\label{sec:heu-group-transfer-common-ancestor-threshold}
Addresses satisfying
\[
\mathrm{hit}(a) \ge 2
\]
are reported as common ancestors, indicating repeated upstream co-occurrence across
multiple cross-chain transfers.
If no such address exists for a given destination transfer, group tracing naturally
degenerates to the single-transfer setting.

\section{The LOCARD Architecture}
\label{sec:architecture}

\begin{figure}
    \centering
    \includegraphics[width=\linewidth]{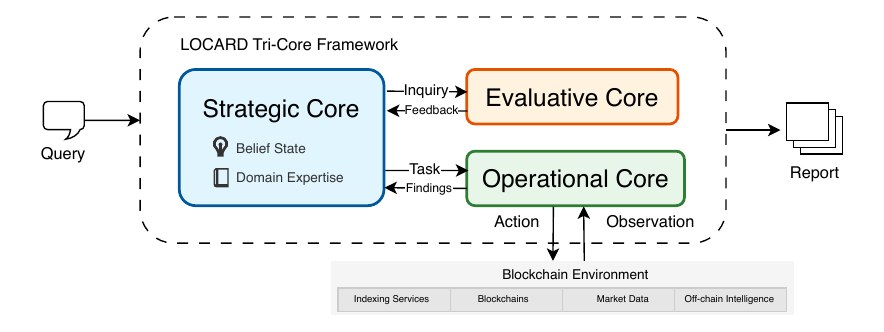}
    \caption{LOCARD's Tri-Core Architecture}
    \label{fig:tri-core}
\end{figure}

We propose LOCARD, an agentic framework designed to automate the cognitive complexity of blockchain forensics. To handle the non-linear nature of cross-chain tracing, LOCARD adopts a \textit{Tri-Core Cognitive Architecture} (Figure~\ref{fig:tri-core}). This design implements a multi-agent system that decouples high-level strategic reasoning from low-level execution and validation.

\subsection{The Tri-Core Cognitive Architecture}
\label{sec:tri_core}

The framework is organized into three distinct functional cores, each responsible for a complementary aspect of forensic investigation.
Rather than forming a linear pipeline, LOCARD coordinates these cores through iterative interaction, enabling adaptive and stateful investigation.

\paragraph{The Strategic Core}
Maintains the structured belief state and governs high-level investigative decisions.

\begin{itemize}
    \item {Belief-Based Reasoning.}
    The Strategic Core maintains and reasons over a structured belief state that captures accumulated findings, unresolved hypotheses, and investigative constraints.
    This belief state enables consistent reasoning across iterative investigation steps.

    \item {Expertise-Guided Strategic Control.}
    Leveraging forensic expertise, the Strategic Core governs high-level strategic decisions, such as when to expand exploration, when to request additional evidence from the Operational Core, and when to invoke validation through the Evaluative Core.
\end{itemize}

\paragraph{The Operational Core}
Executes tool-based evidence acquisition against the blockchain environment.

\begin{itemize}
    \item {Tool Abstraction.}
    The Operational Core encapsulates heterogeneous blockchain data sources, such as RPC nodes and blockchain explorers, into standardized and auditable tool interfaces.

    \item {Iterative Task Execution.}
    It executes the \textit{Task Briefs} issued by the Strategic Core, performing iterative tool invocations as required to traverse transaction graphs and retrieve on-chain evidence.
    In our instantiation, this execution layer is realized as a ReAct-style agent~\cite{yao2023reactsynergizingreasoningacting}, but the framework itself does not mandate an agentic implementation.
\end{itemize}

\paragraph{The Evaluative Core}
Validates and assesses evidences before conclusions are incorporated into the global belief state.

\begin{itemize}
    \item {Evidence Validation.}
    The Evaluative Core performs structural and semantic validation on the \textit{Findings}, checking their internal consistency and contextual plausibility.
    For example, it verifies whether a candidate bridge event is temporally aligned with the corresponding source transaction.

    \item {Evidence Assessment.}
    For validated findings, the Evaluative Core applies domain-specific forensic heuristics (Section~\ref{sec:heuristics}) to assess the plausibility of candidate evidence associations and assign confidence scores.
\end{itemize}

\paragraph{Dynamic Orchestration}
The tri-core interaction is coordinated through a continuous Perception--Reasoning--Action (PRA) loop by the Strategic Core, which governs the progression of the investigation across iterations.
Rather than being confined to a single component, this loop emerges from the coordinated roles of the three cores and is orchestrated by the Strategic Core.

\begin{enumerate}
    \item {Perception.}
    At each iteration, the Strategic Core assimilates new inputs, including raw findings produced by the Operational Core and assessment feedback generated by the Evaluative Core.

    \item {Reasoning.}
    These inputs are integrated into the structured belief state (Section~\ref{sec:structured_belief}), transforming observations and evaluations into coherent forensic context.

    \item {Action.}
    Guided by the updated belief state, the Strategic Core determines the next strategic action, such as issuing additional retrieval tasks, requesting further evaluation, or terminating the investigation.
\end{enumerate}

\subsection{State-Aware Reflection with Structured Belief}
\label{sec:structured_belief}

Large Language Models (LLMs) notoriously suffer from hallucinations and reasoning inconsistencies \cite{wei2023chainofthoughtpromptingelicitsreasoning, huang2023survey}. In complex forensic tasks, these limitations often manifest as unstable behaviors, such as skipping critical verification steps or deviating from the standard operating procedure (SOP) defined by expert heuristics.

To mitigate these risks without sacrificing autonomy, LOCARD implements a \textit{Structured Belief State}, denoted as $B_t$. Unlike unstructured memory, $B_t$ is modeled as a state vector $B_t \in \{0, 1\}^N$, where each dimension represents the completion status of a specific SOP milestone (e.g., \textit{Transfer Identified}, \textit{Validation Passed}).

\paragraph{State-Constraint Reasoning}
The belief state serves as a grounding anchor for the Strategic Core. By explicitly tracking the boolean status of investigation milestones, LOCARD facilitates \textit{contextual action space pruning}. When the agent perceives a state element as \texttt{True}, it implicitly excludes the associated retrieval actions from its future decision boundaries.

\paragraph{State-Aware Reflection}
Crucially, the transition of the belief state ($B_t \rightarrow B_{t+1}$) is not a hard-coded trigger but a decision made by the agent based on execution feedback. The Strategic Core reviews findings returned by the Operational Core, assesses whether they suffice to complete the current sub-task, and only flips the corresponding bit upon acceptance.
This design effectively enforces procedural consistency, preventing common behavioral anomalies such as \textit{premature step-skipping}, \textit{redundant data retrieval}, and the tendency to hallucinate arithmetic or data relationships instead of invoking necessary computational tools.

\section{Instantiation for Cross-Chain Tracing}

\subsection{Workflow Instantiation}
\label{sec:workflow}

\begin{figure}
    \centering
    \includegraphics[width=\linewidth]{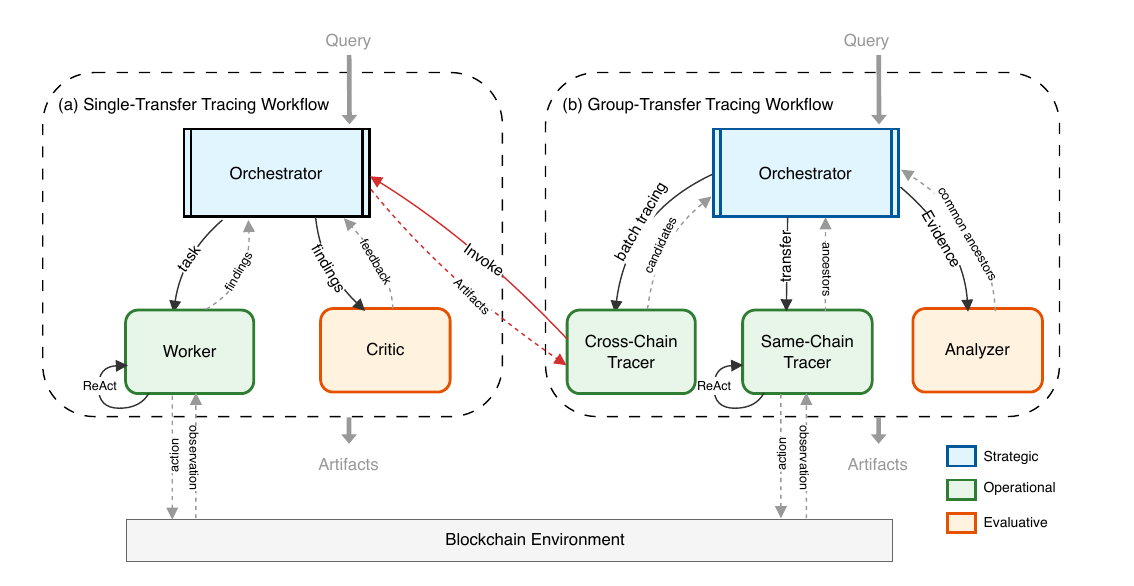}
    \caption{A Workflow Instantiation of LOCARD for Transaction Tracing. (a) Single-transfer cross-chain tracing workflow. (b) Group-transfer cross-chain tracing workflow.}
    \label{fig:workflow}
\end{figure}

To demonstrate the efficacy of the LOCARD framework, we apply and instantiate the architecture to handle the two forensic tasks defined in Section~\ref{sec:tasks}. We implement two specialized workflows (Figure~\ref{fig:workflow}), where agents autonomously navigate the investigation space guided by the heuristics proposed in Section~\ref{sec:heuristics}.

\subsubsection{Single-Transfer Cross-Chain Tracing}
\label{sec:single_instance}

This workflow instantiates the atomic tracing task using three collaborating agents, each aligned with a core in the LOCARD architecture.

\paragraph{The Orchestrator.}
Acting as the strategic core, it governs task decomposition, dispatches the Worker or Critic based on the current state, and updates the state after each feedback round.

\paragraph{The Worker.}
Acting as the operational core, it executes evidence collection through the tool interfaces, including \textit{GetTxByHash}, \textit{SearchTransferByConditions}, and \textit{LookupHistoricalPrice}.
\begin{itemize}
    \item \textit{ReAct Execution:} Iteratively invokes tools in a thought-action-observation loop.
    \item \textit{Feedback Loop:} Reports missing evidence or execution gaps back to the Orchestrator for replanning.
\end{itemize}

\paragraph{The Critic.}
Acting as the evaluative core, it validates the collected evidence and sends feedback signals to the Orchestrator when necessary.

\begin{itemize}
    \item \textit{Integrity Validation:} Checks structural consistency over the findings, such as temporal alignment and chain-level plausibility.
    \item \textit{Likelihood Assessment:} Filters out candidates that violate economic constraints (Eq.~\eqref{eq:heu-value-validation}) and scores the remaining links.

\end{itemize}

\subsubsection{Group-Transfer Tracing}
\label{sec:group_instance}

This workflow handles the Group-Transfer Tracing tasks defined in Section~\ref{sec:group-task} by equipping agentic system with heuristics described in Section~\ref{sec:heu-group-transfer}.

\paragraph{The Orchestrator.} The Orchestrator centrally coordinates three specialized functional modules to reconstruct the shared funding structure. 

\paragraph{The Cross-Chain Tracer.}
This tracer handles the batch cross-chain tracing requests. It dispatches single-transfer tracing cases to multiple instances of the Single-Transfer Workflow (Section~\ref{sec:single_instance}). This design leverages the system's atomic tracing capability to process the disjoint targets concurrently, outputting a set of candidate source transactions.

\paragraph{The Same-Chain Tracer.}
For each identified cross-chain candidate, this module performs upstream graph traversal. It uses chain-specific transfer graph exploration tools to retrieve the predecessor set $\Pred_H(\tau)$ under the same-chain upstream tracing heuristic in Section~\ref{sec:heu-group-transfer}, thereby reconstructing the local funding history on the source chain.

\paragraph{The Analyzer.}
Once the funding graphs are reconstructed, the Analyzer aggregates the resulting evidence. It operationalizes the address-level co-occurrence voting logic by computing the intersections of ancestor sets across the parallel traces, following the common-ancestor voting heuristic in Section~\ref{sec:heu-group-transfer}. Addresses satisfying the consensus threshold ($\mathrm{hit}(a) \ge 2$) are then identified as candidate shared entities controlling the Sybil cluster.

\subsection{Heuristic Evaluation}
\label{sec:scoring}

We implement a lightweight heuristic scoring model to instantiate the decision logic for single-transfer tracing. Note that this scoring module serves primarily to demonstrate the framework's viability and is not claimed to be optimally robust.
The process involves a strict filtration step followed by a confidence computation for surviving candidates.

\subsubsection{Constraints and Metrics}

\paragraph{Filtration} We first discard invalid candidates $\tau'$ that violate causality (i.e., $\ts(\tau') > \ts(\tau^\ast)$) or economic constraints (e.g., negative or excessive implied fees).

\paragraph{Time Proximity ($S_{\text{time}}$).} We model the likelihood of a link decaying exponentially with latency:
\begin{equation}
    S_{\text{time}} = \exp\left( - \frac{\ts(\tau^\ast) - \ts(\tau')}{\lambda} \right)
\end{equation}
where $\lambda$ (e.g., 300s) represents the typical bridge processing window. This metric penalizes significant time gaps which are indicative of unrelated events.

\paragraph{Amount Feasibility ($S_{\text{amt}}$).} To accommodate price volatility without penalizing legitimate arbitrage, we evaluate the width of the \textit{feasible fee rate range} $[r_{\min}, r_{\max}]$ derived from oracle prices with a buffer $\delta$:
\begin{equation}
    S_{\text{amt}} = \frac{r_{\max} - r_{\min}}{\mathcal{R}_{\text{norm}}}
\end{equation}
A wider feasible range implies the candidate is robust to exchange rate uncertainties, whereas a narrow range suggests a coincidental numeric fit.

\paragraph{Aggregation.} The final confidence score is a weighted combination:
\begin{equation}
    S_{\text{final}} = \frac{w_t S_{\text{time}} + w_a S_{\text{amt}}}{w_t + w_a}
\end{equation}
We assign higher weight to timing ($w_t > w_a$) as temporal violations are stronger indicators of falsehood than amount mismatches, which can be conflated by market noise.

\section{The Thor25 Dataset}
\label{sec:dataset}

To evaluate the feasibility and effectiveness of our agentic framework, we construct and release \textit{Thor25}, a comprehensive cross-chain dataset derived from THORChain, a decentralized liquidity network enabling native asset swaps. Unlike existing benchmarks that rely on wrapped tokens or single-chain data, Thor25 captures native layer-1 activities across Bitcoin (BTC), Ethereum (ETH), Dogecoin (DOGE), and Litecoin (LTC) throughout the entire year of 2025. All data is directly derived from THORChain official swap execution records, rather than inferred by tracing heuristics. 

\subsection{Data Acquisition and Tiered Construction}
The raw dataset comprises 151,461 successful cross-chain records. Each record encapsulates a complete swap workflow: an inbound transaction on the source chain, liquidity execution, and an outbound transaction on the destination chain. To facilitate different research needs, we structure the dataset into three tiers:

\begin{itemize}
    \item Thor25 (Raw): The complete set of 151.5k records, preserving the natural distribution of real-world cross-chain traffic, including long-tail low-value transfers and network congestion delays.
    
    \item {Thor25HF (High-Value, Fast):} A refined subset designed to isolate economically significant behaviors and exclude noise. We filter records based on two criteria:
    \begin{enumerate}
        \item {Value Threshold:} We impose dynamic thresholds for source assets (e.g., $\geq 0.09$ BTC, $\geq 1.9$ ETH) to ensure capital significance.
        \item {Temporal Constraint:} We retain only swaps completed within 30 minutes to model low-latency, high-efficiency cross-chain maneuvers.
    \end{enumerate}
    This filtering yields 20,235 high-quality records, constituting a clean corpus for cross-chain pattern analysis.
    
    \item {Thor25HF-Mini:} A balanced random sampling of 1,200 records (100 per pair) derived from Thor25HF. Given the computational complexity of agentic reasoning, this subset serves as the primary testbed for validating our framework's workflow in Section~\ref{sec:experiments-single}. It provides a representative yet resource-efficient environment to demonstrate the feasibility of agent-based forensics.
\end{itemize}

\subsection{Forensic Ground Truth: The Bybit Incident}
A critical contribution of Thor25 is the inclusion of real-world illicit fund flows identified during the Bybit hack in March 2025. Our analysis of the dataset reveals a distinct traffic anomaly: a sudden surge in ETH$\to$BTC volume between March 1 and March 3, coinciding with reported laundering activities.

Within Thor25, we have annotated a subset of traces linked to one of the exploiter addresses\footnote{Reported exploiter: 0xA5A023E052243b7cce34Cbd4ba20180e8Dea6Ad6}. By reconstructing the transaction graph, we identified distinct \textit{Sybil cluster patterns}, where multiple satellite addresses originate from common ancestors. {In this release, we present identified ancestors up to a maximum depth of 3 (Depth 1-3). We explicitly note that this represents a partial snapshot of the laundering network, which likely extends beyond our current search radius.}

While identifying the complete laundering network remains an open challenge due to the complexity of obfuscation techniques, these annotated traces provide a rare and valuable cross-chain ground truth. They serve as a challenging ground truth to validate the detection of sophisticated laundering groups. We release this labeled subset to encourage community contribution toward a more transparent blockchain forensics benchmark.

\section{Experiments}
\label{sec:experiments}


\begin{table}[!t]
\caption{Benchmark results on Thor25HF-mini for the single-transfer cross-chain tracing task.}
\label{tab:crosschain_12pairs_metrics}
\centering
\small
\resizebox{\columnwidth}{!}{%
\begin{tabular}{llrrrrrrr}
\toprule
\textbf{Pair} & \textbf{Method} & \textbf{Recall} & \textbf{Hit@1} & \textbf{Hit@3} & \textbf{Hit@5} & \textbf{Hit@10} & \textbf{Hit@20} & \textbf{Hit@50} \\
\midrule

BTC$\rightarrow$ETH
 & LOCARD     & 100 & 15 & 40 & 57 & 72 & 91 & 100 \\
\rowcolor{lightgray}
 & Heuristic  & 100 & 15 & 40 & 57 & 72 & 91 & 100 \\

BTC$\rightarrow$DOGE
 & LOCARD     & 97 & 15 & 35 & 47 & 67 & 87 & 96 \\
\rowcolor{lightgray}
 & Heuristic  & 98 & 15 & 35 & 47 & 67 & 87 & 97 \\

BTC$\rightarrow$LTC
 & LOCARD     & 99 & 16 & 32 & 49 & 71 & 93 & 98 \\
\rowcolor{lightgray}
 & Heuristic  & 99 & 16 & 32 & 49 & 71 & 93 & 98 \\

ETH$\rightarrow$BTC
 & LOCARD     & 96 & 4 & 12 & 24 & 43 & 66 & 93 \\
\rowcolor{lightgray}
 & Heuristic  & 97 & 4 & 12 & 24 & 43 & 66 & 93 \\

ETH$\rightarrow$DOGE
 & LOCARD     & 94 & 19 & 48 & 63 & 77 & 84 & 92 \\
\rowcolor{lightgray}
 & Heuristic  & 94 & 19 & 48 & 64 & 78 & 84 & 92 \\

ETH$\rightarrow$LTC
 & LOCARD     & 98 & 31 & 60 & 67 & 84 & 95 & 98 \\
\rowcolor{lightgray}
 & Heuristic  & 98 & 31 & 60 & 67 & 84 & 95 & 98 \\

DOGE$\rightarrow$BTC
 & LOCARD     & 100 & 38 & 75 & 94 & 98 & 100 & 100 \\
\rowcolor{lightgray}
 & Heuristic  & 100 & 39 & 75 & 95 & 98 & 100 & 100 \\

DOGE$\rightarrow$ETH
 & LOCARD     & 97 & 62 & 88 & 93 & 95 & 96 & 97 \\
\rowcolor{lightgray}
 & Heuristic  & 98 & 62 & 89 & 94 & 96 & 97 & 98 \\

DOGE$\rightarrow$LTC
 & LOCARD     & 99 & 52 & 89 & 94 & 98 & 99 & 99 \\
\rowcolor{lightgray}
 & Heuristic  & 99 & 52 & 90 & 94 & 98 & 99 & 99 \\

LTC$\rightarrow$BTC
 & LOCARD     & 96 & 8 & 17 & 30 & 63 & 83 & 94 \\
\rowcolor{lightgray}
 & Heuristic  & 98 & 8 & 17 & 30 & 63 & 85 & 96 \\

LTC$\rightarrow$ETH
 & LOCARD     & 96 & 18 & 47 & 69 & 86 & 93 & 96 \\
\rowcolor{lightgray}
 & Heuristic  & 96 & 18 & 47 & 69 & 86 & 93 & 96 \\

LTC$\rightarrow$DOGE
 & LOCARD     & 98 & 27 & 54 & 68 & 80 & 93 & 98 \\
\rowcolor{lightgray}
 & Heuristic  & 98 & 27 & 54 & 68 & 80 & 93 & 98 \\

\bottomrule
\end{tabular}%
}
\end{table}

\subsection{Experimental Setup}

We instantiate the proposed workflow as a multi-agent system orchestrated via LangGraph~\cite{langgraph}, with all agents powered by OpenAI's GPT-4o~\cite{gpt4o_systemcard}. 
Our evaluation leverages the Thor25 benchmark (Section~\ref{sec:dataset}) and combines with a real-world investigation case study (Bybit hack):

\begin{itemize}
    \item {Single-transfer Tracing:} We employ the \textit{Thor25HF-Mini} subset to quantitatively evaluate the framework's baseline tracking performance. This compact dataset balances statistical representativeness with the computational costs of agentic execution. To contextualize the reliability of LOCARD, we additionally implement a heuristic baseline that directly executes the deterministic tracing rules described in Section~\ref{sec:heu-single-transfer}.
    
    \item {Group-transfer Tracing:} We conduct a qualitative case study using the Bybit hack data. This task focuses on a representative Sybil cluster to validate the framework's capacity for basic graph analysis and pattern recognition in real-world illicit flows.
\end{itemize}

\subsection{Single-transfer Tracing Experiment Results}
\label{sec:experiments-single}

\paragraph{Benchmark Results}
Table~\ref{tab:crosschain_12pairs_metrics} summarizes results across 12 heterogeneous cross-chain pairs. LOCARD closely matches the deterministic heuristic baseline in all directions, achieving recall $\geq$93\% and perfect recall in several paths (e.g., BTC$\to$ETH and DOGE$\to$BTC), indicating that the agent framework can faithfully execute forensic tracing heuristics over large on-chain search spaces.
For candidate ranking, Hit@1 varies across pairs due to the intentionally simple scoring rule used in this study, but Hit@50 exceeds 90\% for all directions. Overall, LOCARD narrows the search space to a ranked shortlist of high-probability candidates for downstream analyst validation, rather than asserting definitive links.

\paragraph{Operational Cost and Runtime}
In our experiments, each trace costs approximately \textit{\$0.20} on average, primarily driven by LLM inference tokens, with a runtime of about 1--2 minutes and around 22 LLM calls per trace.
Given that Thor25HF focuses on economically significant transfers (e.g., $\geq$0.09 BTC or $\geq$1.9 ETH), this overhead remains practical for high-value single-transfer forensic tracing, especially considering the high recall rate ($\geq$93\%).
As the framework is model-agnostic, it can further benefit from newer, lower-cost models.

\subsection{Group-transfer Tracing Experiment: A Case Study}
\label{sec:experiments-group}

\paragraph{Bybit Hack Money Flow Scenario}
To evaluate the framework's capability in handling complex obfuscation patterns, we construct a qualitative case study based on the real-world Bybit hack incident. The challenge involves a "fan-out" money laundering pattern: the attacker splits funds from a single Ethereum entity into multiple smaller clusters, bridges them across THORChain, and disperses them into distinct Bitcoin addresses to evade detection. We query the agent with a cluster of 5 disparate Bitcoin transactions (identified as the laundering leaves) and task it with a reverse-lookup objective: to autonomously hypothesize and identify the common upstream entity on Ethereum without any prior knowledge of the transaction graph topology.

\begin{figure}
    \centering
    \includegraphics[width=\linewidth]{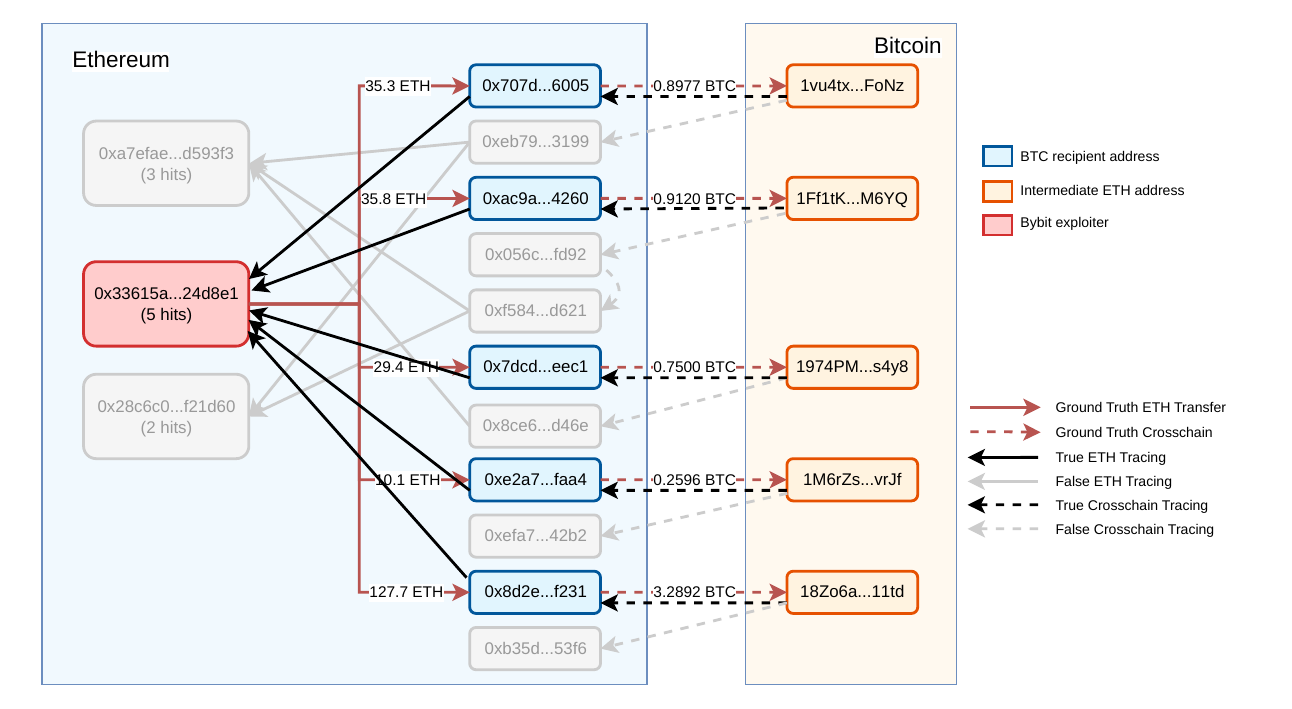}
    \caption{LOCARD reconstructs a subflow of Bybit hack money laundering. (Only partial false tracing results are presented for simplicity)}
    \label{fig:bybit_traceflow}
\end{figure}

\paragraph{Analysis of Reasoning and Results}
The agent successfully reconstructed the cross-chain graph, pinpointing the Ethereum address \texttt{0x3361...} as the unique \textit{Common Ancestor} for all 5 target transactions (see Figure~\ref{fig:bybit_traceflow} for visualization). The agent's reasoning trajectory reveals that it did not merely trace individual paths linearly; instead, it performed dynamic graph intersection analysis. It identified that despite the divergence in execution times and output addresses on Bitcoin, all 5 traces converged to a single source on Ethereum with a \textit{100\% hit rate (5/5 convergence)}. This result validates LOCARD's ability to perform high-level forensic reasoning—detecting Sybil-like structures and correlating fragmented heterogeneous events back to a single root actor.

\section{Discussion}

Our experiments primarily validate faithful heuristic execution under an agentic workflow, rather than superiority over deterministic scripts in the current THORChain setting.

This work demonstrates the feasibility of Agentic Blockchain Forensics in a structured tracing setting where deterministic heuristics already perform strongly. In this setting, LOCARD closely matches the heuristic baseline and replaces a fixed tracing pipeline with an autonomous investigative workflow. This direction is increasingly timely given the rapid recent progress and adoption of agentic systems in real-world software workflows.

More importantly, the value of agentic forensics lies not only in reproducing existing heuristics, but in providing an extensible interface for composing and scaling forensic procedures in broader investigative settings. A natural next step is to extend LOCARD beyond the current THORChain-centered setting through federated specialist agents that operate across heterogeneous chains, bridge protocols, and privacy-oriented environments.

The current scoring component is intentionally simple, and improving candidate scoring remains important for higher ranking precision in larger and more ambiguous search spaces. More adversarial settings, such as dust-style noise injection in low-value traces, also warrant future robustness evaluation.

\section{Conclusion}

This work introduces Agentic Blockchain Forensics (ABF), to our knowledge the first paradigm that frames blockchain investigation as an agentic, evidence-driven process rather than a static tracing pipeline. We instantiate this paradigm through LOCARD, a tri-core framework with structured belief state designed to capture the rigor and balance the exploration-exploitation tension of forensic investigation.

Across benchmark tracing tasks and a real-world laundering case study, LOCARD shows that an agentic system can faithfully execute forensic heuristics over complex cross-chain evidence. We hope this work serves as an initial foundation for future research on ABF in blockchain and beyond.

\section*{Acknowledgment}

Special thanks to Han Yu, Guofeng Yu, and Jing Xiang for their love and support.
For my son.
In memory of my grandmother, Baozhen Tan.

\bibliographystyle{IEEEtran}
\bibliography{ref}

\end{document}